\documentclass[%
 reprint,
 amsmath,amssymb,
 aps,
prb,
]{revtex4-2}

\usepackage{graphicx,color}
\usepackage{dcolumn}
\usepackage{bm}
\usepackage{soul}

\begin{document}

\preprint{APS/123-QED}

\title{    Armchair carbon nanotube on Pt and hBN/Pt: from strong metallic contact to coherent spin transport regime}

\author{Marcin Kurpas}
\email{marcin.kurpas@us.edu.pl}
\affiliation{%
 Institute of Physics, University of Silesia in Katowice, 41-500 Chorzów, Poland 
}%
\author{Marko Milivojevi\'c}
\affiliation{Institute of Informatics, Slovak Academy of Sciences, 84507 Bratislava, Slovakia}
\affiliation {Faculty of Physics, University of Belgrade, 11001 Belgrade, Serbia}
\author{Martin Gmitra}
\affiliation{Institute of Physics, Pavol Jozef \v{S}af\'{a}rik University in Ko\v{s}ice, 04001 Ko\v{s}ice, Slovakia}
\affiliation{Institute of Experimental Physics, Slovak Academy of Sciences, 04001 Ko\v{s}ice, Slovakia}
\date{\today}

\begin{abstract}
We study spin-orbit proximity effects in an armchair (4,4) carbon nanotube on the Pt(111) surface. By employing first-principles calculations, we show that the Dirac cone of the metallic nanotube is altered due to strong hybridization with the Pt substrate. Inserting a monolayer hexagonal boron nitride (hBN) between the nanotube and the substrate limits the hybridization effects leading to recovering the Dirac cone. The Dirac bands display asymmetric spin splitting, 0.7\,meV for the right movers and 1.7\,meV for the left movers at the K valley, due to the proximity to the Pt substrate. 
We find that the Dirac states exhibit almost perfect spin polarization, transverse to the nanotube axis and to the stacking direction, forming a proper condition for charge-to-spin conversion with coherent spin transport in the nanotube. 
 We propose an effective Hamiltonian describing the proximity-induced effects on the Dirac electrons and their spin texture.

\end{abstract}

\maketitle
\section{\label{sec:intro}Introduction}

Carbon nanotubes (CNTs) were synthesized years before graphene and have been intensively studied for many years \cite{Iijima1991,Laird2015}. 
Being essentially rolled graphene sheets, carbon nanotubes display much broader electronic properties than their flat counterpart.
While graphene is a semimetal with a vanishing density of states at the Fermi level, carbon nanotubes can be either semiconducting or metallic \cite{saitobook1998}. 
It makes them perfect candidates for applications in digital electronics, since the semiconducting band gap is intrinsic and does not require interference with the crystal structure \cite{Tans1998,zhou2000,Mueller2010,Hills2019}.
Much stronger than in graphene spin-orbit coupling (SOC) \cite{Gmitra2009,Sichau2019}  and the absence of hyperfine coupling   \cite{Ando2000,Chico2004,Kuemmeth2008,Jeong2009,Izumida2009,Steele2013} makes CNTs also attractive for applications in spintronics and quantum technologies \cite{Laird2013,Laird2015,Cubaynes2019}.

The source of enhancement of SOC in CNTs comes from the lattice curvature, which  enables coupling of $\sigma-\pi$ orbitals, forbidden in flat graphene by symmetry 
 \cite{Ando2000,huertas2006,Jeong2009,Izumida2009}. Therefore, stronger SOC is expected for small-diameter nanotubes due to large curvature. In armchair CNTs, which should be nominally metallic, SOC opens an orbital gap at the $K$-point proportional to the inverse diameter $\Delta_{\text{so}}\sim 1/d$ \cite{Izumida2009,zhou2009}. According to first principles calculations  $\Delta_{\text{so}}$ can reach a few meV \cite{zhou2009}.
 On the other hand, spin splittings in Dirac cone bands induced by a transverse electric field are much smaller than spin-orbital gaps,  on the order of tens $\rm \mu eV$  per V/nm \cite{Klinovaja2011a}. Such small splittings can lead to experimental difficulties in accessing individual spin states of armchair nanotubes, for instance, to realize helical modes \cite{klinovajaRPL2011}.
 Alternative methods of increasing energy separation of spin states are needed in order to fully exploit the potential of carbon nanotubes in spintronics and quantum technologies.

The success of exploiting proximity effects to modify the electronic properties of graphene and recent progress in the fabrication of one-dimensional (1D) van der Waals heterostructures has brought a revival of the interest in carbon nanotubes \cite{Xiang2020,Jadwiszczak2022,Matsushita2023}. While the fabrication of concentric nanotube heterostructures still seems to be very challenging, nanotubes on surfaces of bulk materials, in particular metals, are very common in the experiment  \cite{Tans1997,Bockrath1997,Mann2003}.  
Moreover, recent technology allows for controlled growth and fabrication of aligned arrays of wafer-scale carbon nanotubes
\cite{Zhang2017,He2020,Liu2020CNT} overcoming the main obstacles to applying nanotubes in microelectronics and spintronics and opening excellent prospects for practical applications of nanotubes.

So far, the interaction of carbon nanotubes with bulk surfaces of two-dimensional (2D) materials has been studied mainly in the context of orbital effects, charge transfer, and contact formation or proximity-induced superconductivity \cite{Morpurgo1999,MAITI20047,okada2005,Park2005,Zhu2006,Zhu2006,Meng2007,Casterman2009,Shin2009,Nishidate2010,Shao2014,Mackus2017,Bauml2021}. 
Not much attention has been put on SOC and spin physics in such systems.
Since the cylindrical geometry makes the electronic properties of carbon nanotubes much different from graphene's, one can expect that it will also affect proximity effects.
In particular, finite curvature and diameter of the tube can create specific conditions for the crystal potential at the interface \cite{Hasegawa2011} and related spin-orbit effects.

In this paper, we use first-principles calculations to discuss proximity spin-orbit effects in a (4,4) armchair single-wall carbon nanotube in contact with Pt(111) surface. We chose platinum due to the strong spin-orbit coupling and its widespread use as electrical contacts to carbon nanotubes \cite{Cao2005,Franklin2014,Mackus2017}. 
We show that in contrast to flat graphene/Pt heterostructure, in which the relatively weak graphene-Pt interaction leaves Dirac cone bands to a large extent intact \cite{Sutter2009, Khomyakov2009},
states of a small carbon nanotube hybridize strongly with Pt states, and the characteristic Dirac cone can no longer be identified in the electronic band structure. 
To restore the Dirac bands, we place a single layer of hexagonal boron nitride (hBN) between the CNT and the Pt slab. It suppresses direct chemical bonding and turns the system into a proximity regime in which the Dirac cone of the nanotube is preserved.
The proximity-induced SOC leads to spin splitting of the Dirac bands at the meV level and results in an asymmetric energy spectrum, similar to that of an armchair carbon nanotube in coexisting electric and magnetic fields \cite{Klinovaja2011a}. Bands of the nanotube are almost perfectly spin-polarized and realize spin-valley locking due to the opposite polarization direction for the two valleys. 
We describe the observed effects with an effective four-band model Hamiltonian and identify spin-orbit terms and external fields.

The paper is organized as follows. In Sec. \ref{sec:methods}  we describe methods and technical details of first principles calculations.  In Sec. \ref{sec:results} we present and discuss numerical results and formulate an effective Hamiltonian capturing essential physics. Closing remarks and conclusions are given in Sec. \ref{sec:conclusions}.

\section{\label{sec:methods}Methods}
First-principles calculations were performed using the {\sc Quantum Espresso} software package \cite{QE-2009,QE-2017}.  Ultrasoft pseudopotentials with the Perdew-Burke-Ernzerhof (PBE) \cite{perdew_1996,*perdew_1997}  implementation of the generalized gradient approximation (GGA) exchange-correlation functional was used. The kinetic energy cutoff of the plane wave basis sets was $48\,$Ry for the wave function, and $528$\,Ry for charge density. These values were found to give converged results also for spin-related quantities. Van der Waals interactions were taken into account within the semi-empirical Grimme's DFT-D2 corrections \cite{grimme2006,barone2009}.
A vacuum of 16\,\AA~ in the stacking direction ($x$ in Fig. \ref{fig:cnt_pt}) was introduced to avoid spurious interaction between periodic copies of the system. The dipole-correction \cite{Bengtsson1999} was also included in the calculations. 
Before starting the actual calculations the geometry of the structure was optimized by relaxing internal forces acting on atoms. This step was done separately for non-relativistic and relativistic cases using the quasi-Newton scheme, as implemented in the {\sc Quantum Espresso}, and setting the thresholds $10^{-3}$\,Ry/bohr for force $10^{-4}$\,Ry/bohr for the total energy, respectively. During relaxation, the positions of Pt atoms were frozen in all directions. This constraint was released for a few Pt atoms on the Pt surface lying directly below the CNT in the CNT/Pt heterostructure (Fig. \ref{fig:cnt_pt}). All other atoms were free to move in all directions. 
Self-consistency was achieved with $4\times 1\times 15$ Monkhorst-Pack \cite{MPack} grid.

\section{Results and discussion}\label{sec:results}

\begin{figure}
    \centering
    \includegraphics[width=\columnwidth]{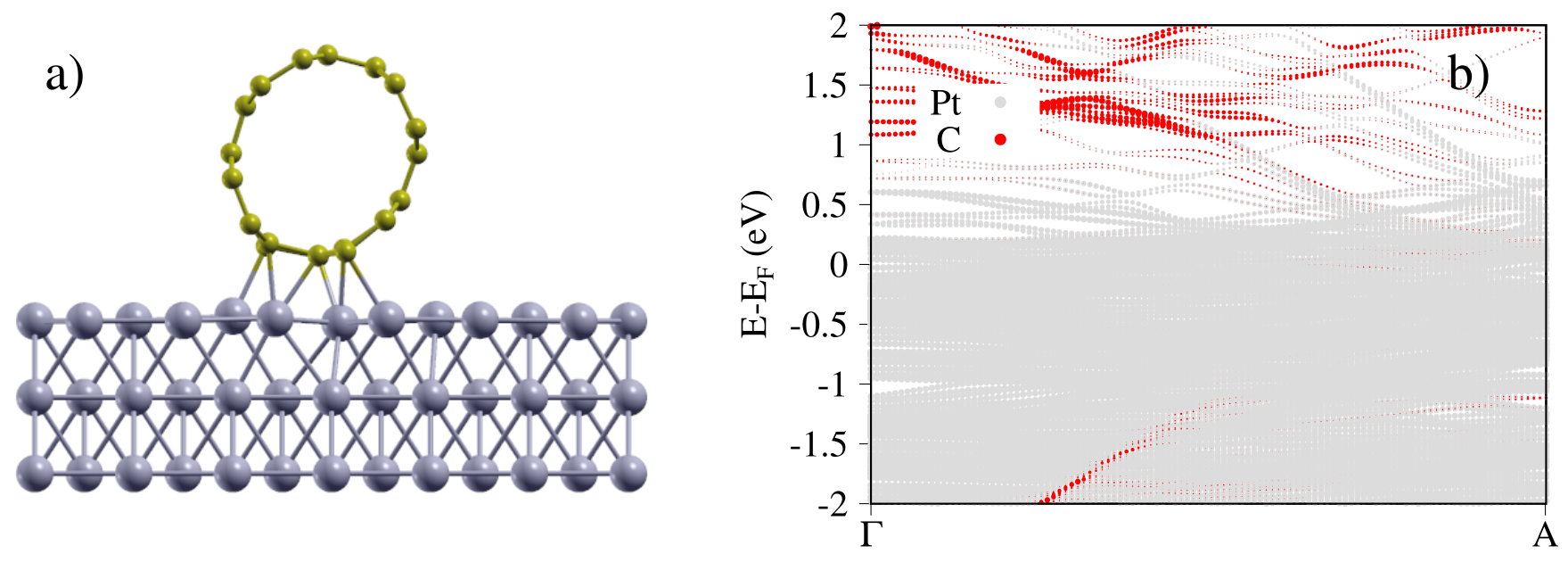}
    \caption{Armchair (4,4) carbon nanotube on Pt (111) surface. (a)~Optimized crystalline structure. The nanotube deformation confirms the formation of a strong chemical bonding. (b)~Atom-resolved relativistic band structure from first principles. Due to strong hybridization, the states of the CNT close to the Fermi level are smeared out by Pt states.}
    \label{fig:cnt_pt}
\end{figure}

We start by discussing the heterostructure made of the metallic (4,4) armchair nanotube on a platinum slab terminated by the (111) surface. 
Previous works discussing nanotubes on Pt predicted the formation of a metallic contact between the nanotube and the surface \cite{MAITI20047,Hasegawa2011}.
In Fig.~\ref{fig:cnt_pt}(a) we show the optimized crystal structure of the heterostructure. A significant slimming of the nanotube with a sizeable difference between vertical and horizontal diameters 0.56\,\AA~is observed. The distance between the tube and Pt surface is 2.1\,\AA, very similar to the 1.98\,\AA~for metallic (10,10) nanotube on Pt \cite{Hasegawa2011}, and 0.3\,\AA~smaller than for the semiconducting (8,0) nanotube \cite{MAITI20047}. The discrepancy between semiconducting and metallic tubes can be explained by a limited possibility of charge redistribution in semiconducting in comparison to metallic systems. 
On the other hand, the smaller curvature of (10,10) CNT does not cause as strong deformation as for (4,4) and (8,0) nanotubes, despite the semiconducting character of the latter. This is related to the suppression of amplitudes of $\pi$ orbitals inside the tube and their amplification outside the tube due to a large curvature in small diameter nanotubes \cite{Hasegawa2011}.

The close distance of the CNT to the surface suggests a strong hybridization of the nanotube and Pt states. Indeed, the atom-resolved band structure shows, see Fig.~\ref{fig:cnt_pt}(b), that the states of the nanotube are completely smeared out by the Pt states, and their contribution to the total charge at the Fermi level is negligible. This confirms that the nanotube forms a good contact with the Pt.

\begin{figure}
    \centering
    \includegraphics[width=0.98\columnwidth]{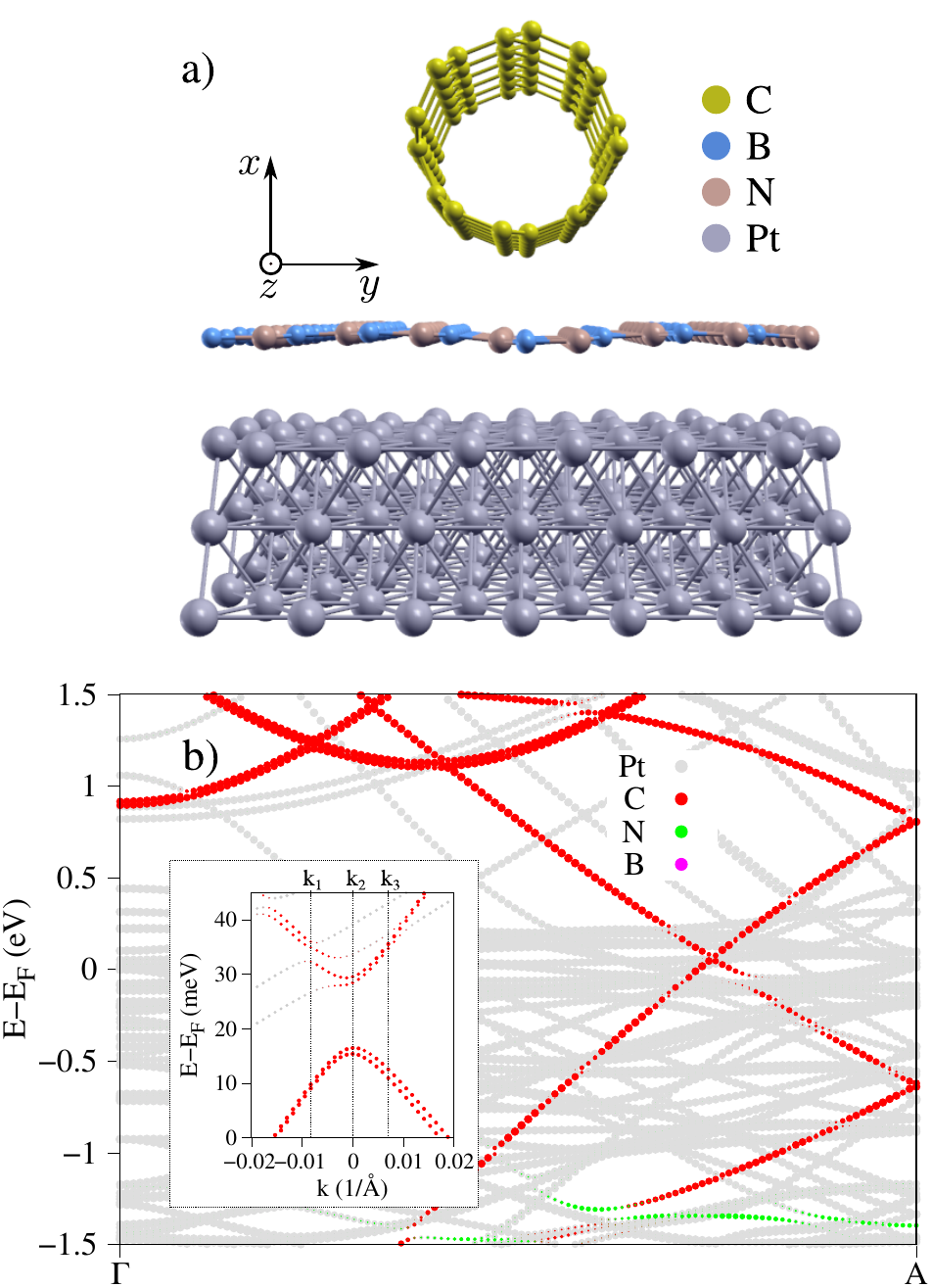}
    \caption{Armchair (4,4) carbon nanotube on hBN/Pt: (a)~optimized structural model of the studied system.  (b)~Calculated non-relativistic atom-resolved band structure. Bands of the nanotube as shown by the red circles, Pt bands with gray, N with green, and B with magenta.
    The inset shows the relativistic band structure around the Dirac cone of the nanotube. Momenta emphasized with the vertical lines are guides for spin splitting analysis.
    \label{fig:bs_cnt_hbn_pt}}
\end{figure}

To limit the hybridization effects, we insert an atomic layer of hBN in between the nanotube and the Pt slab. Since hBN is an insulator, protection of the Dirac cone in the CNT/Pt heterostructure is expected. 
The optimized geometry of the CNT/hBN/Pt heterostructure is shown in Fig.~\ref{fig:bs_cnt_hbn_pt}(a). Now, the nanotube is almost perfectly circular, with only 0.03\,\AA~difference between vertical and horizontal diameters. The distance between the Pt surface and the hBN layer is 3.1\,\AA, and between hBN and the nanotube is 3\,\AA, which are the typical values for van der Waals heterostructures.

Figure \ref{fig:bs_cnt_hbn_pt}(b) shows the calculated atom-resolved electronic band structure. The Dirac bands of the nanotube can be easily identified in the whole range of crystal momenta confirming their effective protection by the hBN monolayer from strong hybridization with Pt states. The valence bands of the tube are free of hybridization effects up to 40\,meV below the band maximum, and should be easily accessible in the experiment. 
In the conduction bands of the nanotube, the hybridization effects are indicated by the anticrossings, see inset in Figure \ref{fig:bs_cnt_hbn_pt}(b), but most of the orbital band character preserves the carbon character. The energy gap between the Dirac cone branches is 12\,meV, and is dominated by the orbital effects due to the presence of the substrate. 
The Fermi level lies 22~meV below the valence band maximum of the CNT  indicating weak hole doping of the nanotube.

Pristine armchair nanotubes are highly symmetric structures \cite{Damnjanovic1999}. Due to the combination of the rotational and vertical mirror symmetry, the spin splitting in armchair nanotubes is forbidden~\cite{lazic18,milivojevic18}.
Thus, without breaking the time-reversal symmetry, bands at any crystal momenta $k$ are spin doublets, even if the spin-orbit coupling is considered.
The presence of the substrate breaks all those symmetries removing the band spin degeneracy. The Kramers doublets  $E_{\sigma, \mathbf{k}}= E_{-\sigma, -\mathbf{k}}$, $\sigma=\lbrace \uparrow, \downarrow\rbrace$ are not affected unless the time-reversal symmetry is broken.

In the inset to Fig.~\ref{fig:bs_cnt_hbn_pt}(b) we show a relativistic band structure in the vicinity of the Dirac point. 
 The spin splitting of the CNT states is asymmetric and equals for the momenta $k_1$ and $k_3$ for the valence bands to 0.73\,meV and 1.7\,meV, respectively. 

The conduction bands of the nanotube in the regions free of hybridization are also split asymmetrically, by 1.5\,meV and 0.8\,meV for $k_2$ and $k_3$ respectively. 
Such large splitting amplitudes can be explained only by the substrate-induced spin-orbit proximity effect. The symmetry breaking of the nanotube caused by the presence of the substrate alone is not able to produce such large splitting values, even even though the intrinsic SOC in Dirac bands of CNT is two orders stronger than for graphene \cite{Gmitra2009,zhou2009}. We have checked, that for a pristine (4,4) CNT at the K point, the splitting energy generated by a transverse external electric field is of about $6\,{\rm\mu eV}$ per V/nm, almost the same as in graphene \cite{Gmitra2009}.

Similarly, we relate the observed asymmetry to the presence of the substrate-induced interface potential and SOC effects. 
As it was shown in an analogous study of (4,4) carbon nanotube on a monolayer bismuthene, the asymmetry of Dirac cone bands strongly depends on the local atomic environment created by the substrate~\cite{Kurpas2023}. By modifying this environment, for instance, by shifting the nanotube on the substrate, one can control the topology and spin texture of the Dirac cone bands. Similar effects of asymmetry have been found in 2D mixed-lattice heterostructures, whose incompatible symmetries can trigger novel types of spin textures in the target materials~\cite{MGK+23,MGK+24,MGK++24}.

\begin{figure}
    \centering
    \includegraphics[width=\columnwidth]{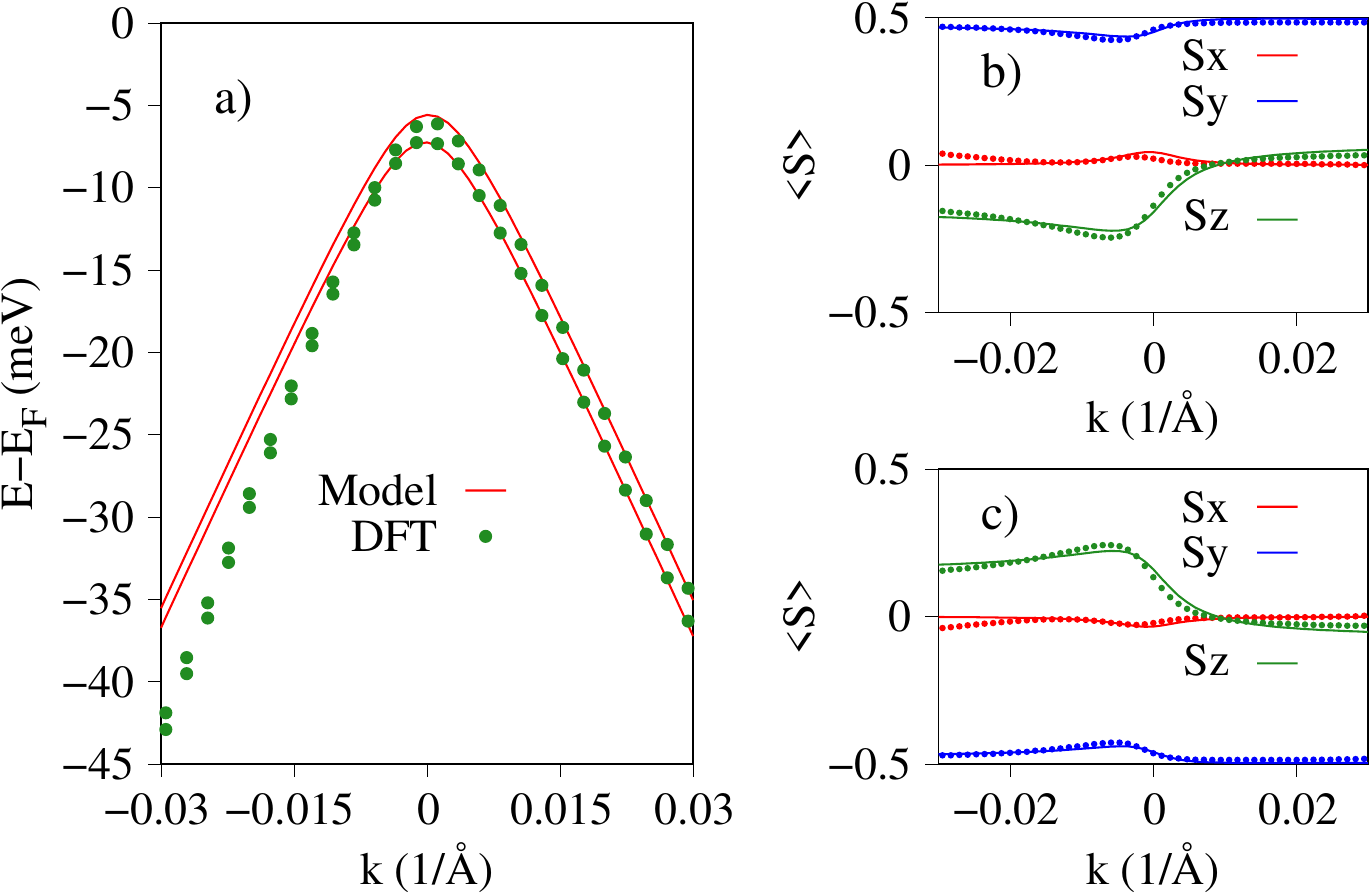}
    \caption{Comparison of the effective model (lines) with first principles results (dots): (a)~valence bands of the nanotube centered around the Dirac point; (b)~spin expectation values for the upper split band, and (c)~for the lower spin split band shown in (a). In the whole range of $k$ the bands are almost perfectly spin-polarized along the $y$-direction (perpendicularly to the CNT axis). }
    \label{fig:model_dft} 
\end{figure}

An important observation is that the Dirac bands are almost perfectly spin-polarized in the $y$ direction, see Fig.~\ref{fig:model_dft} (b)-(c), that is, perpendicularly to the nanotube axis and to the stacking direction of the heterostructure. A small $S_z$ component is visible for negative crystal momenta, which may lead to a slight asymmetry for spin scattering for the left and right movers, but in general, the spin polarization is momentum independent which should result in long spin coherence times due to strongly suppressed Dyakonov-Perel \cite{dyakonov1971,dyakonov1971R} spin relaxation mechanism \cite{holleitner2006,schliemann2017}. The non-zero $S_z$ component may be possibly reduced by fine-tuning the position of the nanotube on the substrate \cite{Kurpas2023}. A similar anisotropic band structure exists at the $K'$ valley but with opposite spin polarization due to time reversal symmetry.

 \textit{Effective Hamiltonian}.
 To give a more intuitive picture of the physics underlying the observed effects we develop an effective Hamiltonian capturing essential features of the first principles band structure and spin texture. 
 We start with the Hamiltonian of the pristine armchair nanotube with intrinsic SOC reading \cite{Izumida2009,Klinovaja2011a}
 \begin{equation}
     H_{\rm cnt} = \tau \hbar v_F k \sigma_2 + \alpha S_z \sigma_1 .
     \label{eq:hHcnt}
 \end{equation}
 Here $v_F$ is the Fermi velocity, $k$ is crystal momentum along the tube, $\sigma_i$ are the Pauli matrices acting on the sublattice space,
 $S_z$ is the  spin one-half operator with eigenvalues $\pm 1$, and $\tau$ is the valley index, $\tau=1$ ($\tau=-1$) for the $K$ ($K^{'}$) valley. 
 This Hamiltonian gives spin degenerate Dirac cone bands and opens a spin-orbital gap at the $K$-point of size  2$\alpha$. 
 
Next, we add a staggered, sublattice-odd, potential $\Delta_{\rm st}$ diagonal in spin subspace, which opens an orbital gap at the $K$-point 
\begin{equation}
    H_{\rm st} = \Delta_{\rm st} \sigma_3.
    \label{eq:staggered}
\end{equation}
Here, the gap opening mechanism differs from that in non-armchair nanotubes, for which the Hamiltonian contains $\sigma_1$ matrix mixing the sublattices \cite{Izumida2009}. The orbital gap opened by $H_{\rm st}$ is much bigger than the spin-orbital gap. The latter, however, affects the electron spin and can not be neglected when discussing spin-orbit effects.

We also consider a Rashba-like Hamiltonian originating from the crystal potential asymmetry in the out-of-plane (\textit{x}) direction 
\begin{equation}
    H_{\rm R} =\tau \alpha_R S_y \sigma_2,
    \label{eq:rashba}
\end{equation}
where $\alpha_R$ is the strength of the Rashba SOC. This term causes a symmetric splitting of the spin degenerate bands and polarizes electron spins in the $y$ direction, similar to the transverse electric field \cite{klinovajaRPL2011}.

Finally, we introduce additional SOC terms triggered by the symmetry-free environment of the studied heterostructure 
\begin{eqnarray}
  H_{\rm eff} &=&     \tau \Omega_1 S_y\sigma_0 + \tau \Omega_2  S_z\sigma_3 + \tau \Omega_3  S_z\sigma_2.   
\label{eq:SOfield}
\end{eqnarray}
The $H_{\rm eff}$ originates from the interface crystal potential but extends physical description not captured in Eqs.~(\ref{eq:staggered}) and (\ref{eq:rashba}). 
The first term in (\ref{eq:SOfield}) describes a sublattice even effective magnetic field in the $y$ direction. It is a time-reversal symmetric analog of an external transverse magnetic field \cite{Klinovaja2011a} leading a spin-dependent shift in $k$ and the asymmetry of spin splitting for left- and right-movers (when $E_x \neq 0)$. 
The second term is an effective, sublattice-odd spin-orbit field polarizing spins along the tube axis (in the $z$ direction). It is analogous to the valley-Zeeman SOC in graphene with broken pseudospin symmetry \cite{gmitra2013,Gmitra_2015} and originates from a sublattice-dependent Kane-Mele SOC term \cite{Kane2005}. The last term in (\ref{eq:SOfield}) also contributes to $S_z$ spin component, but in contrast to $S_z\sigma_3$ it involves hopping between sublattices. 

Parameters of the full Hamiltonian $H = H_{\rm cnt} + H_{\rm st} + H_{\rm R} + H_{\rm eff}$ were found by fitting the energy spectrum and spin textures to the first principles data. Results are shown in Fig.~\ref{fig:model_dft}. The corresponding values of the model parameters are collected in Table \ref{tab:table1}. 
Besides a small discrepancy in the band dispersion in the left part of the Dirac cone resulting from different values of $v_F$ for left and right movers in the first principles band structure, the model perfectly reconstructs DFT results. One can easily notice  (see Table \ref{tab:table1}), that $\Omega_1$ is much bigger than $\Omega_2$ and $\Omega_3$. Thus $E_x$ and $\Omega_1$, being analogs of transverse electric and magnetic fields, have a major impact on low energy electronic properties of the nanotube. 

\begin{table}[b]
\caption{\label{tab:table1} Parameters of the effective model Hamiltonian. All values are in meV if not specified otherwise.}
\begin{ruledtabular}

\begin{tabular}{lcccccr}
{$v_F$\,(m/s)}&
{$\alpha$}&
{$\alpha_R$}&
{$\Delta_{\rm st}$}&
{$\Omega_1$ }&
{$\Omega_2$ }&
{$\Omega_3$ }\\
\colrule 
1.8$\cdot 10^{15}$ & -1.6 &0.27 &6.2 & 0.8 & 0.28& -0.16\\
\end{tabular}
\end{ruledtabular}
\end{table}

\section{Conclusions}\label{sec:conclusions}
We have performed a systematic study of the orbital and spin properties of an armchair carbon nanotube on Pt(111) and hBN/Pt(111) by means of first principles calculations. We have shown, that a strong and destructive to Dirac cone bands hybridization of nanotube and Pt states can be effectively limited by inserting a single-layer hBN spacer. As a result,  a tunnel contact between the nanotube and Pt slab is formed and the system is turned into the spin-orbit proximity regime, in which spin-orbit coupling in the Dirac bands of the nanotube is greatly enhanced, to the meV range. The bands display almost perfect spin polarization creating perfect conditions for coherent spin transport through the nanotube.

\begin{acknowledgments}
M.K. acknowledges support from the Interdisciplinary Centre for Mathematical and Computational Modelling (ICM), University of Warsaw (UW), within grant no. G83-27 and the financial support from the National Center for Research and Development (NCBR) under the V4-Japan project BGapEng V4-JAPAN/2/46/BGapEng/2022.
M.M. acknowledges the financial support
provided by the Ministry of Education, Science, and Technological Development of the Republic of Serbia. This project has received funding from the European Union's Horizon 2020 Research and Innovation Programme under the Programme SASPRO 2 COFUND Marie Sklodowska-Curie grant agreement No. 945478.
M.G.~acknowledges financial support provided by the Slovak Research and Development Agency provided under Contract No. APVV SK-CZ-RD-21-0114
and Slovak Academy of Sciences project IMPULZ IM-2021-42 and project FLAG ERA JTC 2021 2DSOTECH.
\end{acknowledgments}

\section{References}

\bibliography{bibliography}

\end{document}